\documentclass[11pt]{article}
\usepackage{moriond}
\usepackage{graphicx}

\usepackage{subfig}
\usepackage{comment}
\usepackage{caption}
\usepackage{xspace}
\usepackage{cite}

\def\Journal#1#2#3#4{{#1} {\bf #2}, #3 (#4)}

\def\NPB{{\em Nucl. Phys.} B}

\def\PRL{\em Phys. Rev. Lett.}

\def\mus{\ensuremath{\mu s}\xspace}
\def\dt{\ensuremath{\Delta t}\xspace}
\begin{document}
\vspace*{4cm}
\title{TOWARD A SUB-PPM MEASUREMENT OF THE FERMI CONSTANT}

\author{ D.M. WEBBER }

\address{Department of Physics, University of Illinois at Urbana-Champaign\\
1110 West Green Street, Urbana, IL, USA\\
{\em for the MuLan Collaboration} \footnote{
S. Battu, R.M. Carey, D. Chitwood, P.T. Debevec,
S. Dhamija, W. Earle, A. Gafarov, F.E. Gray,
K. Giovanetti, T. Gorringe, Z. Hartwig, D.W. Hertzog,
B. Johnson, P. Kammel, B. Kiburg, S. Kizilgul,
J. Kunkle, B. Lauss, I. Logashenko, K.R. Lynch, R. McNabb,
J.P. Miller, F. Mulhauser, C.J.G. Onderwater, Q.Peng, J. Phillips,
S. Rath, B.L. Roberts, V. Tishchenko, D.M.Webber, P. Winter, B.Wolfe}}
\maketitle\abstracts{
The Fermi constant, $G_F,$ describes the strength of the weak force and is determined most precisely from the mean life of the 
positive muon, $\tau_\mu$. Advances in theory have reduced the theoretical uncertainty
on $G_F$ as calculated from $\tau_\mu$ to a few tenths of a part per million (ppm). The remaining uncertainty
on $G_F$ is entirely experimental, and is dominated by the uncertainty on $\tau_\mu$.
The MuLan experiment is designed to measure the muon lifetime to part-per-million precision,
a better-than twenty-fold improvement over the previous generation of experiments.
In 2007, we reported an 
intermediate result, $\tau_\mu=2.197013(24)$ $\mathrm{\mu s}$ (11 ppm), which is in
excellent agreement with the previous world average. 
This mean life was measured using
a pulsed surface muon beam stopped in a ferromagnetic target, surrounded by a symmetric
scintillator detector array. 
Since this intermediate measurement, the detector was instrumented with waveform digitizers,
the muon beam rate and beam extinction were increased, and two data sets were acquired on
different targets, each containing over $10^{12}$ muon decays. These data will lead to a
new determination of $G_F$ to better than a part per million. 
}

The Standard Model (SM) has provided an excellent description of fundamental particles and their interactions since its introduction in the early 1970's. 
The SM description of the weak force requires three input parameters;  the most precisely determined are the fine structure constant, $\alpha$, the Fermi constant, $G_F$, and the mass of the Z-boson, $M_Z$.
The Fermi constant, $G_F$, is determined most precisely from a measurement of the positive muon lifetime, $\tau_\mu$,
via the formula 

\begin{equation}
  \frac{1}{\tau_\mu} = \frac{G_F^2 m_\mu^5}{192 \pi^3} \left(1+\Delta q\right),
  \label{eqn:RS Form}
\end{equation}
where $\Delta q$ includes corrections from phase space, QED, hadronic and tau loops.  In this parameterization, weak and non-weak effects factorize and all weak corrections are included in $G_F$ as $\Delta r$,
\begin{equation}
  %\frac{G_F}{\sqrt{2}}=\frac{g_W^2}{8M_W^2}\left(1+\Sigma_i r_i\right).
  \frac{G_F}{\sqrt{2}}=\frac{g_W^2}{8M_W^2}\left(1+\Delta r \right),
\end{equation}
where $g_W$ is the electroweak gauge coupling.  The calculation of the second-order QED corrections to the determination of $G_F$ from $\tau_\mu$ reduce the theoretical uncertainty on $\Delta q$ from $\sim 30$~ppm to less than 0.3~ppm \cite{vanritbergen-2000-564}, making the uncertainty on $\tau_\mu$ the dominant uncertainty on $G_F$.
This theoretical progress motivated the current generation of muon lifetime experiments. 
%Given the positive muon's importance, it should be measured as accurately as possible. 
%The positive muon lifetime is also used as a benchmark for muon capture experiments, such as muon capture on the proton (MuCap) \cite{MuCap_proposal}, and muon capture on the deuteron (MuSun) \cite{MuSun_proposal}.

The Muon Lifetime Analysis (MuLan) experiment is designed to measure the muon lifetime to part-per-million (ppm) precision using a time-structured muon beam and a symmetric, segmented detector surrounding a central muon-stopping target.  We presented an intermediate result, $\tau_\mu=2.197013(24)$ $\mathrm{\mu s}$ (11 ppm), based on data collected in 2004 \cite{mulan2004}.  Since that measurement, the experimental electronics were upgraded from discriminators to waveform digitizers, and two high-statistics datasets were collected on different stopping targets in 2006 and 2007, respectively.

%The MuLan experiment's beamline and detector are optimized for a precise measurement of the muon lifetime.  
The experiment runs in a multi-muon mode where several muon decays are measured simultaneously.  Muons are provided by a 10 MHz dc beam modulated by a fast-switching, 25-kV electrostatic kicker \cite{Barnes:2004}.  The time-structure imposed by the kicker has a 5-\mus ``accumulation period,'' during which muons are accumulated in a stopping-target, followed by a 22-\mus ``measurement period'' to observe the decay of the accumulated muons.  During the kicker-on measurement period, the beam rate is reduced by a factor of $\sim1000.$ Approximately 50 muons arrive and stop in the target during the accumulation period, and 20 survive until the beginning of the measurement period.  An instability in the kicker voltage could affect the observed muon lifetime, but the uncertainty on $\tau_\mu$ from kicker instabilities has been determined to be less than 0.2~ppm for the 2006 dataset and less than 0.07~ppm for the 2007 dataset. 

%One systematic concern is muon spin precession.  
The positive muons accepted by the beam optics are produced from pion decay at rest and have negative helicity.  This helicity is preserved as the muons are transported along the beamline and stop in the target.  A slow precession or relaxation of the average muon spin, combined with a possible asymmetry in detector acceptance, could introduce a perturbation on the observed muon lifetime.  To prevent an unobserved slow spin precession or relaxation from perturbing the observed muon lifetime, two targets were chosen to dephase or visibly precess the average muon spin.  
In 2006 a ferromagnetic target with $\sim0.5$~T internal magnetic field, called Arnokrome-3 (AK3) \cite{AK3}, dephased the average muon spin during the accumulation period.
In 2007, a disk of quartz crystal caused $\sim90\%$ of the muons to form muonium, a hydrogen-like bound state composed of one muon and one electron.  In the spin-0 state, the spin of the muon and the spin of the electron are rapidly exchanged, effectively nulling the muon spin effects.  In the spin-1 state, the spins of the electron and muon are locked together by the hyperfine interaction, and precess $\sim103\times$ faster than a bare muon \cite{hughes_wu}.  
An array of permanent magnets provide a 130-Gauss field to visibly precess the remaining muons.
The magnetic field in both target configurations was oriented transverse to the beam and initial muon-spin direction.
%A high percentage of the muons that stop in the quartz form muonium
%, and the remainder were spin-precessed in a 130-Gauss magnetic field produced by an array of permanent magnets.  
%These two datasets were analyzed independently.
%\footnote{Both targets could be opened for beam monitoring using a wire chamber downstream of the detector}.

\begin{figure}[t]
  \centering
  \subfloat[]{
  \includegraphics[height=0.35\textheight]{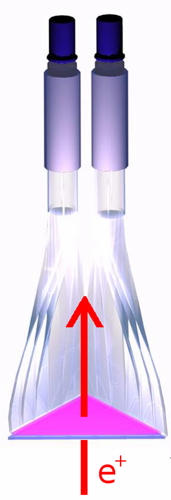}
  }
  \hspace{36pt}
  \subfloat[ ]{
  \includegraphics[height=0.35\textheight]{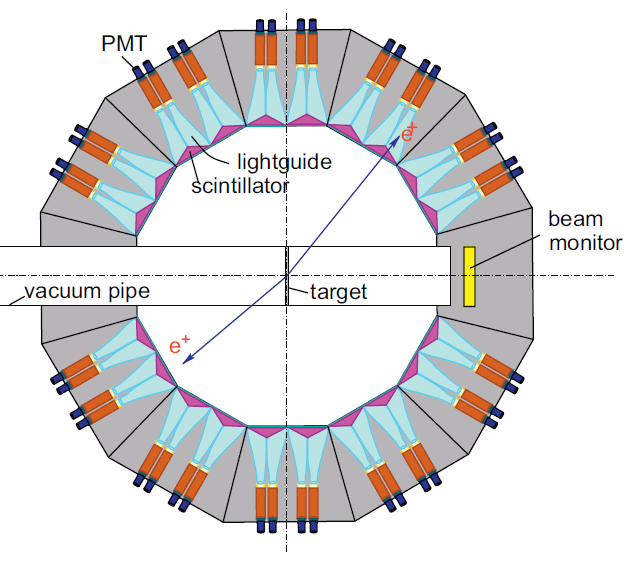}
  }
  \caption{Illustrations of the detector.  (a) A through-going positron will pass through the inner and outer triangular scintillator in a tile pair.  (b) A cartoon shows the location of the target inside the vacuum beampipe and centered in the detector. %adapted from \cite{tishchenko2008data}.
  }
  \label{fig:ball_diagram}
\end{figure}

Muon decay $\mu^+\rightarrow e^+\ \overline{\nu}_\mu\ \nu_e$ occurs with $\sim100\%$ probability.  Positrons from muon decay in the stopping target are observed by the MuLan detector.
The detector is composed of a truncated icosahedron (soccer ball) configuration of triangular scintillator tile pairs which symmetrically surrounds the target
(Fig. \ref{fig:ball_diagram}).  
This detector segmentation results in  low individual tile pair count rate, $\sim 0.1$ hit per tile pair per beam cycle.
The detector configuration contains 20 hexagonal detector housings, each containing 6 tile pairs, and 10 pentagonal detector housings, each containing 5 tile pairs.  
Two pentagonal faces are open, allowing the vacuum pipe to pass through the detector.  The stopping target is suspended in the center of the detector, inside the vacuum pipe.  
The symmetry of the detector largely cancels any asymmetries from residual average muon spin precession or relaxation.  
%The highly segmented detector, with 170 independent tile pairs, results in $\sim0.1$ muon decay per detector per beam beam cycle.

Several systems are required to read out the scintillation light produced in the tile pairs by a through-going positron from muon decay.
Each tile is connected to an adiabatic lightguide and phototube, and 
the analog pulses from the phototubes are digitized by 8-bit waveform digitizers (WFDs).
The WFDs trigger when the analog input rises above threshold, and write out 24 samples ($\sim 53$~ns) of waveform.  
Each WFD board has four analog inputs, assigned to a tile pair and its symmetrically opposite pair.  The 85 WFD boards required for detector readout are distributed in 6 VME crates, and each crate is readout by a dedicated frontend computer.  

A backend computer collects the data and controls the time-structure of the data acquisition (DAQ).  The kicker runs continuously, but the DAQ acquires 5000 beam-cycles at a time, called a segment.  After each segment, the computers read out how much data is available in the WFD FIFO memory, and then triggers the next segment.  Data from the previous segment is readout, compressed, and assembled by the backend while the current segment is being acquired.  This parallel structure of the DAQ allows for data rates up to $\sim40$~MB/s with no decrease in DAQ live-time.
The data acquisition is discussed in more detail elsewhere \cite{tishchenko2008data}.

The digitization frequency for the Waveform Digitizers is provided by a master clock, set to a frequency of $\sim451$~MHz.  During data collection and analysis, the precise clock frequency was concealed to allow for a blind analysis.  During the analysis, times were reported in units of clock ticks (ct), where 1~ct$\approx$2.2~ns. After the analysis was complete, the true clock frequency was revealed.
Different blinding offsets were used in 2006 and 2007, and the datasets were analyzed independently.

\begin{figure}[t]
  \centering
  \subfloat[][]{
  \includegraphics[width=0.45\textwidth]{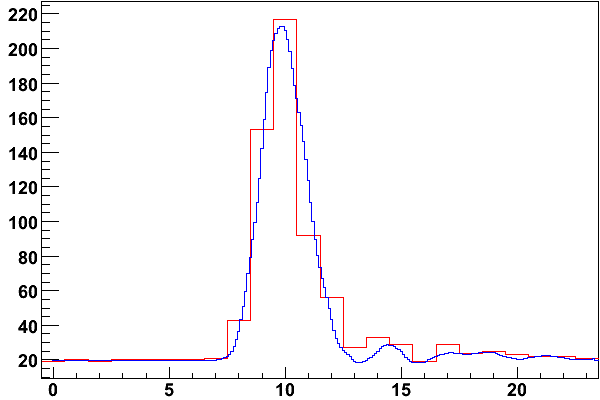}
  \label{fig:pulse_a}
  }
  \subfloat[][]{
  \includegraphics[width=0.45\textwidth]{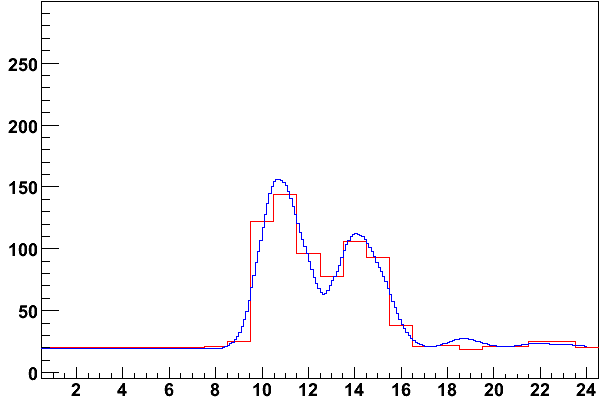}
  \label{fig:pulse_c}
  }
  \caption{Examples of pulses show the resolving power of the fitting code.  When triggered, the WFD records 24 samples ($\sim53$~ns) of waveform data.  The raw data is shown in red and the fitted pulse template is shown in blue.  %\ref{fig:pulse_a} 
  (a) A fit to a normal single-pulse waveform.  Over 90\% of the data is comprised of similar single-pulse waveforms.  %\ref{fig:pulse_b} shows 
  (b) The fitter is able to resolve two pulses if they are separated by 3 or more samples.
  }
  \label{fig:pulses}
\end{figure}

\begin{figure}[ht]
  \centering
  \subfloat[2006]{
  \includegraphics[width=0.47\textwidth]{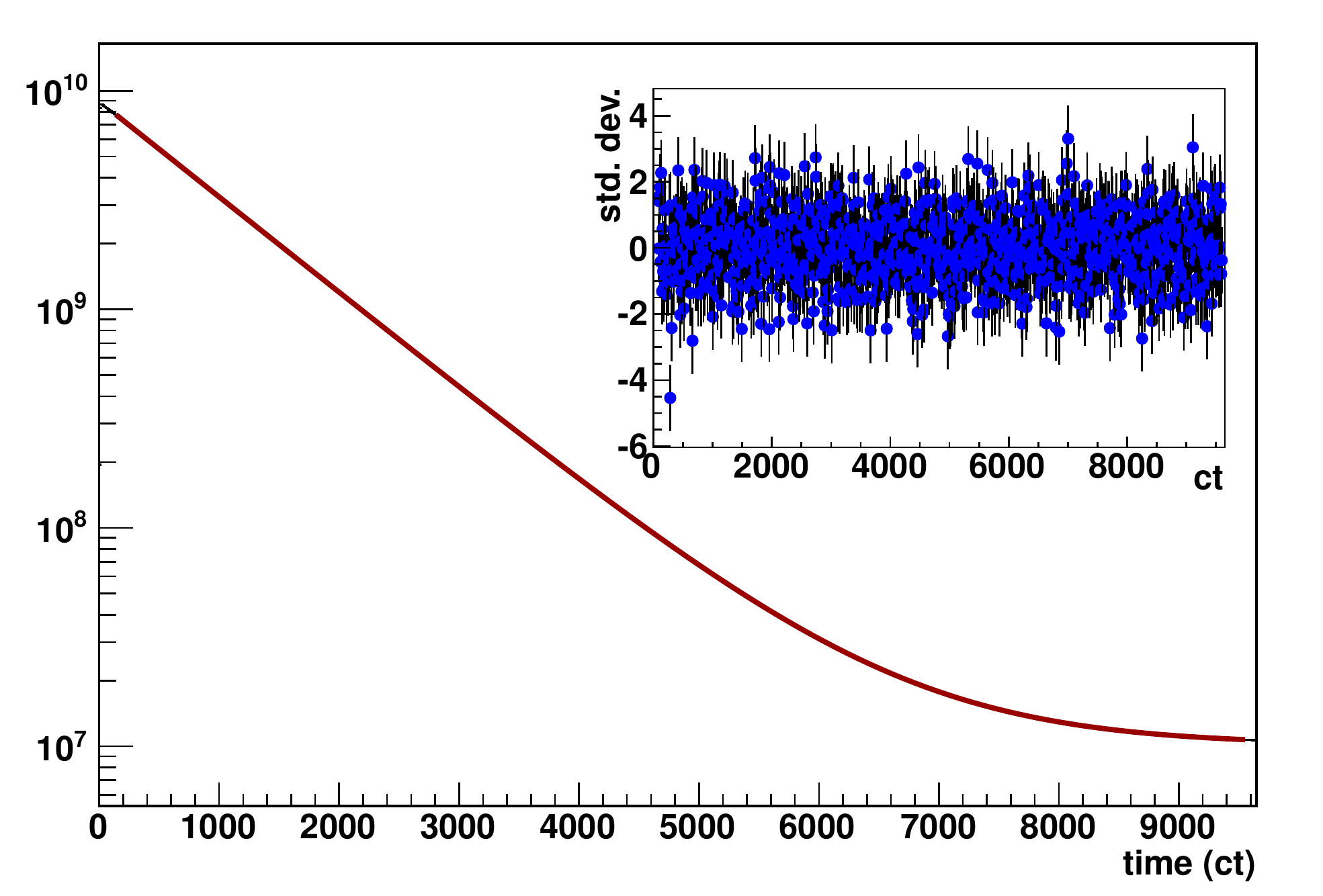}
  \label{fig:lifetime_2006}
  }
  \subfloat[2007]{
  \includegraphics[width=0.47\textwidth]{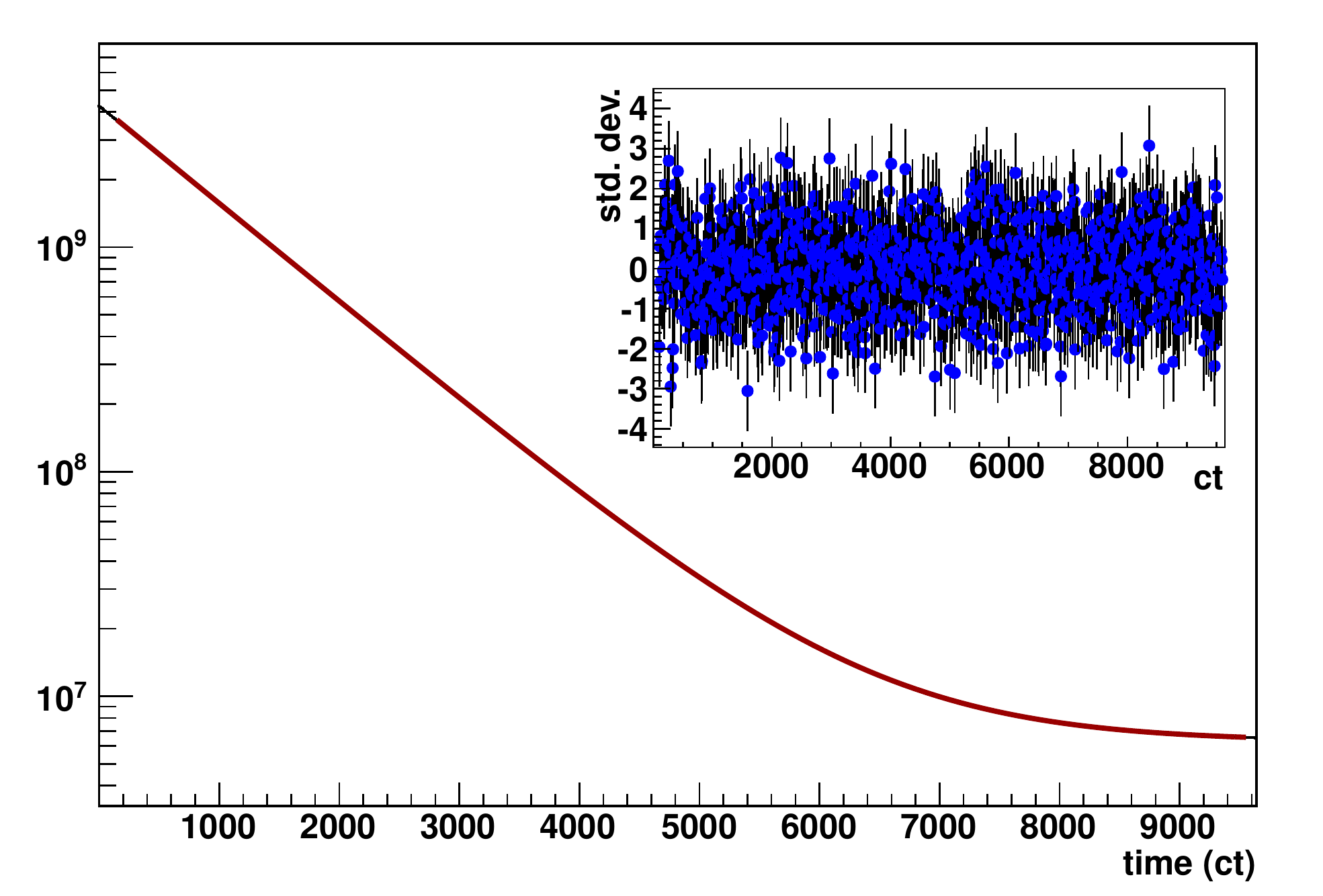}
  \label{fig:lifetime_2007}
  }
  \caption{Lifetime histograms and fit residuals for the 2006 and 2007 datasets.
  }
  \label{fig:lifetimes}
\end{figure}

Overall, $\sim10^{12}$ muon decays were collected on each of the ferromagnetic AK3 and quartz targets, respectively.
The analysis is performed in two stages.  
In the first stage, each waveform is fit using pulse templates to determine the pulse time and height, as shown in figure \ref{fig:pulses}.  In the second stage, coincidences are formed between inner and outer tiles of each pair, and the coincidences are histogrammed vs. time in measurement period.  
%The resulting histograms for each dataset are shown in figure \ref{fig:lifetimes}.  
The histograms are corrected for pileup and for multiple hits from single-source events.  The histograms are then fit with the function 
\begin{equation}
  f(t) = (1 + A S(t)) N_0 e^{-t/\tau_\mu} + B,
  \label{eqn:lifetimeSpline}
\end{equation}
where 
%$\tau_\mu=\tau_\mathrm{secret}(1+R/10^6) $ and 
$A S(t)$ is small correction for a WFD threshold oscillation, 
%of order $10^{-4}$, 
originating in the VME logic fanout, $N_0$ is the number of muon decays at $t=0$, and $B$ is the flat background.  The magnitude of the electronics oscillation correction on the lifetime is 0.60~ppm, and its uncertainty 0.26~ppm.  The lifetime histograms for the two datasets, with fit residuals, are shown in figure \ref{fig:lifetimes}.  

\begin{figure}[ht]
  \centering
  \includegraphics[width=0.8\textwidth]{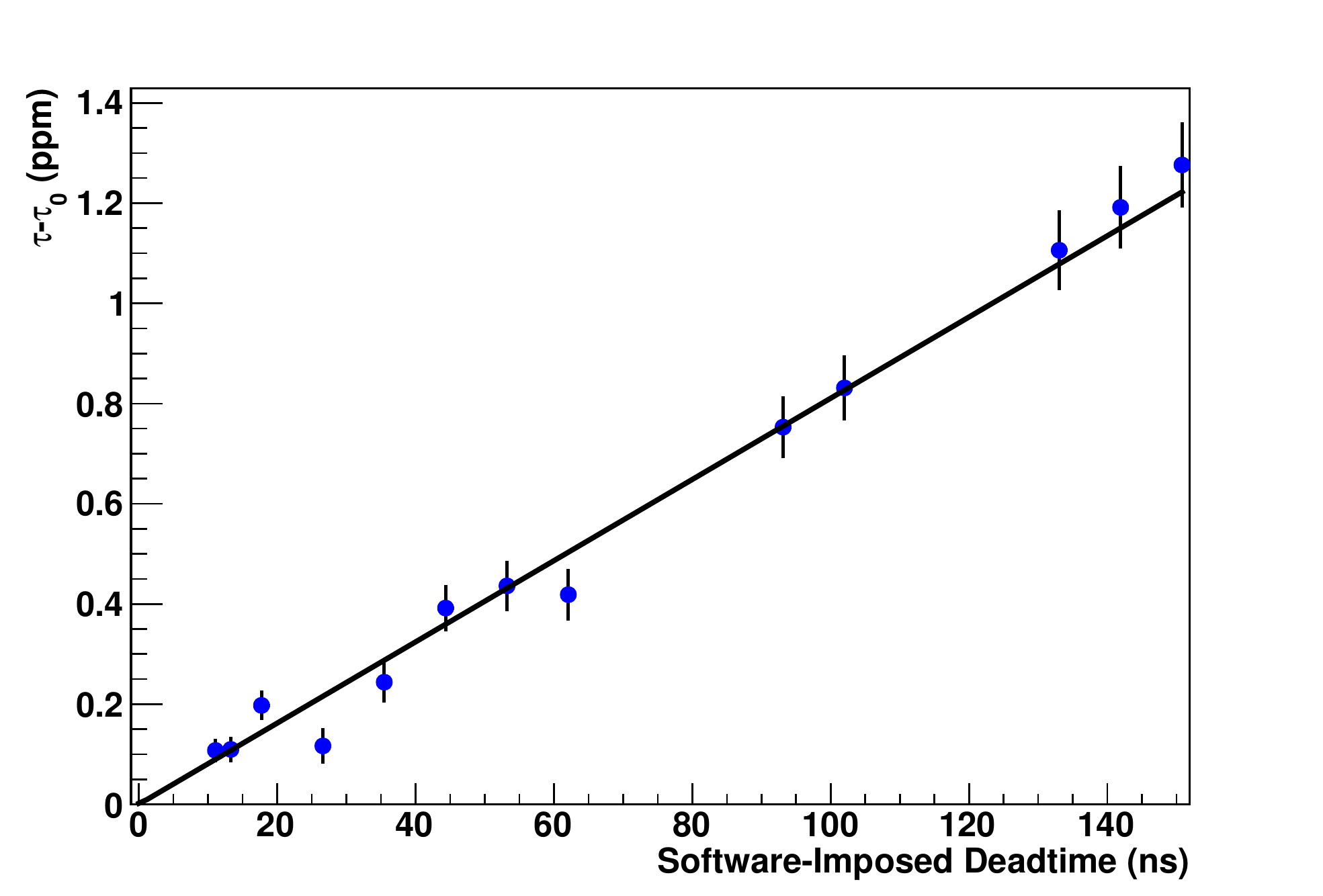}
  \caption{Shift in pileup-reconstructed lifetime vs. software-imposed deadtime for the 2006 dataset.  The uncertainty on each point represents the uncertainty of the statistical pileup correction.  The overall statistical uncertainty of the 2006 dataset is 1.18 ppm.  A small slope of 0.008 ppm per ns deadtime is observed.
  }
  \label{fig:R_vs_ADT}
\end{figure}

Although the WFDs digitize continually, particles passing through a detector close together in time will be incorrectly reconstructed by the analysis, resulting in missed events.
This loss of hits is called pileup.  The pileup is statistically corrected using a shadow-window technique.  To test the fidelity of the corrections, a set of software deadtimes (ADTs) in the range 5-68~samples (11-151~ns) are imposed after a hit at time $t_i$ in beam cycle $j$.
The pileup is corrected by searching the same deadtime interval $t_i$ to $t_i+{\rm ADT}$ in beam cycle $j+1$.  If a ``shadow'' hit is found in this interval, it is added back into the lifetime histogram.
Several different classes of pileup were corrected.  A Monte-Carlo simulation of detector hits shows no dependence between $\tau_\mu$ and deadtime, but a dependency of 0.008 ppm/ns deadtime is observed in the data (Fig. \ref{fig:R_vs_ADT}).  
Several effects that could pull $\tau_\mu$ vs. deadtime have been ruled-out, and the dependency likely corresponds to $\sim0.1\%$ under-correction in the leading-order pileup term.  Under-corrected leading-order pileup introduces a linear dependence between $\tau_\mu$ and ADT.  In this case, a linear extrapolation to zero deadtime gives the correct value for $\tau_\mu$.  Overall, the uncertainty on the pileup corrections and the extrapolation is 0.2~ppm.

\begin{table}[h!]
  \centering
  \caption{Systematic Uncertainties, in units of parts-per-million.  These uncertainties represent preliminary upper bounds on the uncertainties, and will decrease as studies are finalized.  Some systematic uncertainties vary between datasets.  Correlated uncertainties are denoted by a single number, and uncorrelated uncertainties are given with one number for each running year.  Effects vs. time are evaluated vs. time in measurement period, and effects vs. \dt are evaluated vs. time after a prior pulse.}
  \label{tbl:ErrorTable}
  \begin{tabular}{lcc}
    \hline
    Effect                          & \multicolumn{2}{c}{Size (ppm)} \\
    & 2006        &    2007 \\
    \hline
    \hline
    Kicker stability                & 0.22  & 0.07 \\
    Spin precession                 & n/a   & 0.20 \\
    Clock calibration               & \multicolumn{2}{c}{0.03} \\
    Errant muon stops               & \multicolumn{2}{c}{0.10} \\
    Gain stability vs. time         & \multicolumn{2}{c}{0.70} \\
    Gain stability vs. \dt          & \multicolumn{2}{c}{0.27} \\
    Timing stability vs. time       & \multicolumn{2}{c}{0.09} \\
    Timing stability vs. \dt        & \multicolumn{2}{c}{0.08} \\
    Electronics stability vs. time  & \multicolumn{2}{c}{0.26} \\
    Pileup correction               & \multicolumn{2}{c}{0.20} \\
    %Residual polarization & n/a & 0.2 \\
    \hline
    Total systematic                & \multicolumn{2}{c}{0.85} \\
    \hline
    Statistical uncertainty         & 1.18 & 1.7  \\
    \hline
    Total uncertainty                & \multicolumn{2}{c}{1.3} \\
    \hline
    \hline
  \end{tabular}
\end{table}

Other systematic effects are shown in table \ref{tbl:ErrorTable}.  The largest uncertainty at 0.7~ppm is the effect on $\tau_\mu$ from the detector response, or gain, vs. measurement time.  This effect is still under investigation, and the value reported here is an upper limit.  

%Another systematic concern is when hits in the detectors arrive close together in time and are lost due to software-imposed deadtime.  
%Although the beam provides $\sim$20 muons to the detector, the segmented detector to keep the individual detector rates low.
%The high beam rate beam rate is offset by a highly segmented detector, keeping the individual detector rates low. 
%To keep the per-channel rate manageable, The high beam rate is compensated by a highly segmented detector 

\begin{figure}[h]
  \centering
  \includegraphics[height=0.8\textwidth,angle=90]{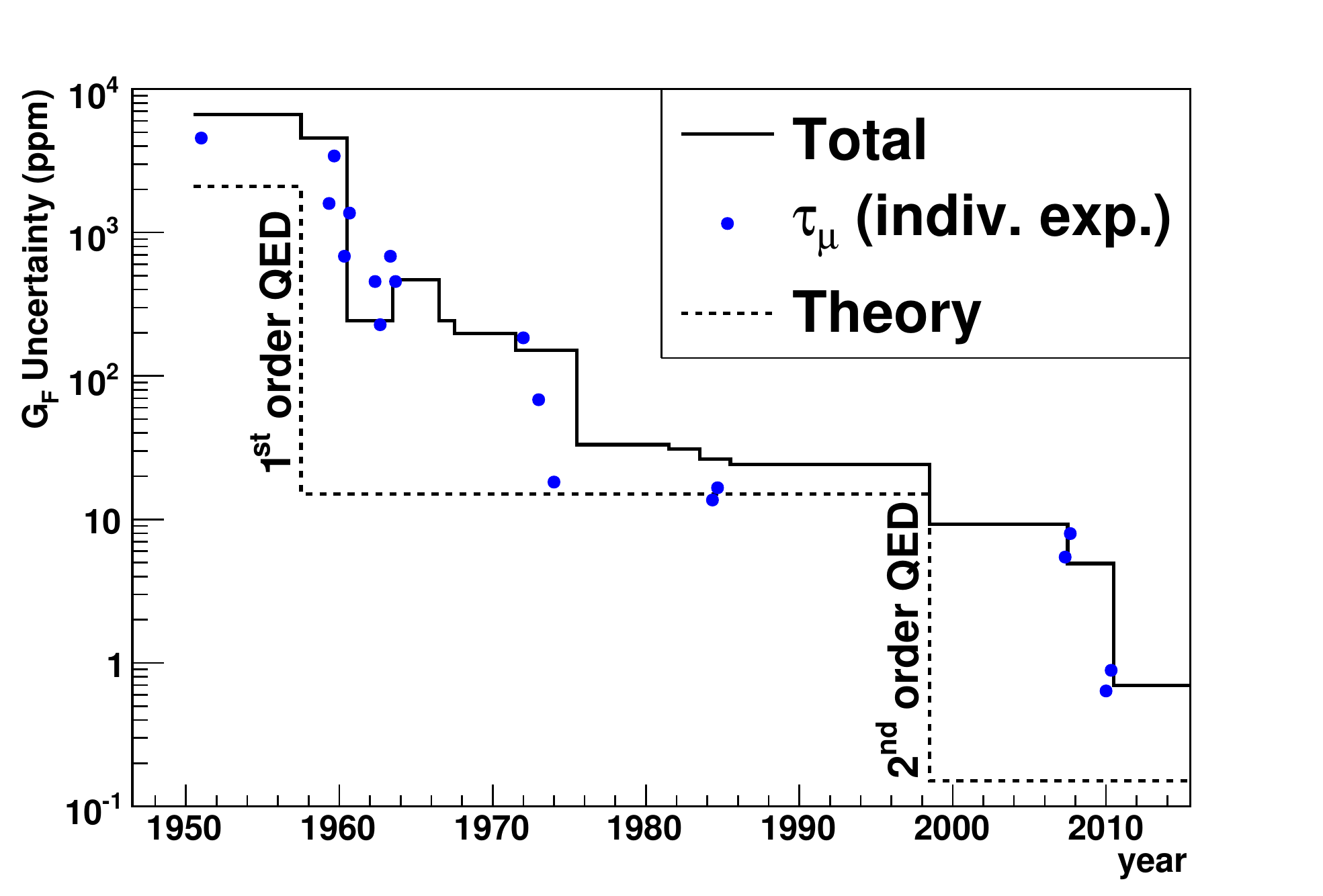} % old root
  \caption{Theoretical and experimental uncertainty on $G_F$ vs. time.  The second-order quantum electrodynamics (QED) corrections to $G_F$ reduced the theoretical uncertainty on $G_F$ from ~30 ppm to ~0.3 ppm, making experimental uncertainty on the muon lifetime the dominant contribution to the uncertainty on $G_F$.}
  \label{fig:precision_vs_time}
\end{figure}

Overall, the MuLan experiment has measured $\tau_\mu$ to ppm-level precision, and will soon publish the final results.
Prior to fully unblinding, a relative unblinding, which mapped the two datasets into a common blinded space, showed the lifetime difference between the two datasets is 0.3~ppm, an excellent agreement.  
Currently, a few systematic uncertainties are undergoing final analysis, and the publication is in preparation.
A plot showing the historical precision of $G_F$ and $\tau_\mu$ is shown in figure \ref{fig:precision_vs_time}, including two points in 2010 representing the $\tau_\mu$ results from our two datasets.  After this measurement, the value for $\tau_\mu$ will have a precision of 1.3~ppm or better, and $G_F$ will have a precision of 0.8~ppm or better.

%A preliminary comparison between our results and recent measurement for the muon lifetime is shown in figure \ref{fig:horizontal}.  

%\begin{figure}[h]
%  \centering
%  \includegraphics[width=0.8\textwidth]{horizontal.pdf}
%  \caption{Recent muon lifetime measurements \cite{balandin1974measurement,giovanetti1984mean,bardin1984new,mulan2004,fast} compared to the average (green line) of our 2006 and 2007 results.}
%  \label{fig:horizontal}
%\end{figure}

\section*{Acknowledgments}
Support 
%for my travel and stipend 
was provided by the National Science Foundation (NSF) under grant NSF PHY 06-01067, and the experiment hardware was largely funded by a special Major Research Instrumentation grant, NSF 00-79735, ``Collaborative Research: The MuLan Project - Development of instrumentation for a new high-precision determination of the Fermi coupling constant.'' 
This work was partially supported by the National Center for Supercomputing Applications under research allocation TG-PHY080015N and development allocation PHY060034N and utilized the ABE and MSS systems.

\end{document}